\documentclass[letterpaper]{article} 
\usepackage{aaai25}  
\usepackage{times}  
\usepackage{helvet}  
\usepackage{courier}  
\usepackage[hyphens]{url}  
\usepackage{graphicx} 
\usepackage{natbib}  
\usepackage{caption} 
\usepackage{algorithm}
\usepackage{algpseudocode}
\usepackage{multirow}
\usepackage{xcolor}
\usepackage{colortbl}
\usepackage{booktabs}
\usepackage{pifont}
\usepackage{amssymb}
\usepackage{enumitem}
\usepackage{amsmath}
\usepackage{listings}

\newif\ifDEBUG

\DEBUGtrue

\ifDEBUG
    \newcommand{\YHL}[1]{\textcolor{blue}{[YHL: #1]}}     
    \newcommand{\NJE}[1]{\textcolor{red}{[NJE: #1]}}     
    \newcommand{\PJJ}[1]{\textcolor{cyan}{[PJJ: #1]}}    
    \newcommand{\NIR}[1]{\textcolor{brown}{[NIR: #1]}}   
    \newcommand{\KNG}[1]{\textcolor{orange}{[KNG: #1]}}  
    \newcommand{\BSHC}[1]{\textcolor{purple}{[BSHC: #1]}} 
    \newcommand{\KYY}[1]{\textcolor{green}{[KYY: #1]}} 
    \newcommand{\TPN}[1]{\textcolor{olive}{[TPN: #1]}} 
\else
    \newcommand{\YHL}[1]{}
    \newcommand{\NJE}[1]{}
    \newcommand{\PJJ}[1]{}
    \newcommand{\NIR}[1]{}
    \newcommand{\KNG}[1]{}
    \newcommand{\BSHC}[1]{}
    \newcommand{\KYY}[1]{} 
    \newcommand{\TPN}[1]{} 
\fi

\usepackage{newfloat}
\usepackage{listings}

\DeclareCaptionStyle{ruled}{labelfont=normalfont,labelsep=colon,strut=off} 
\floatstyle{ruled}
\newfloat{listing}{tb}{lst}{}
\floatname{listing}{Listing}
\title{Detecting Music Performance Errors with Transformers}

\newcommand{\Polytune}[0]{\textit{Polytune} }

\author{
    Benjamin Shiue-Hal Chou, 
    Purvish Jajal, 
    Nicholas John Eliopoulos, 
    Tim Nadolsky,\\
    Cheng-Yun Yang, 
    Nikita Ravi, 
    James C. Davis, 
    Kristen Yeon-Ji Yun, 
    Yung-Hsiang Lu
}
\affiliations{
    School of Electrical and Computer Engineering, Purdue University, West Lafayette, IN, USA, 47907\\
    \{chou150, pjajal, neliopou, tnadolsk, yang2316, ravi30, davisjam, yun98, yunglu\}@purdue.edu
}
\begin{document}

\maketitle

\begin{abstract}

Beginner musicians often struggle to identify specific errors in their performances, such as playing incorrect notes or rhythms. 
There are two limitations in existing tools for music error detection: 
(1) Existing approaches rely on automatic alignment; therefore, they are prone to errors caused by small deviations between alignment targets.; 
(2) There is a lack of sufficient data to train music error detection models, resulting in over-reliance on heuristics. 
To address (1), we propose a novel transformer model, \textit{Polytune}, that takes audio inputs and outputs annotated music scores. 
This model can be trained end-to-end to implicitly align and compare performance audio with music scores through latent space representations. 
To address (2), we present a novel data generation technique capable of creating large-scale synthetic music error datasets. 
Our approach achieves a 64.1\% average Error Detection F1 score, improving upon prior work by 40 percentage points across 14 instruments. 
Additionally, compared with existing transcription methods repurposed for music error detection, our model can handle multiple instruments. Our source code and datasets are available at \texttt{https://github.com/ben2002chou/Polytune}.

\end{abstract}

\section{Introduction}
\label{intro}
Beginner musicians often need help identifying errors in their performance. 
For example, novice musicians may struggle with sight reading or miss notes due to a lack of muscle memory.
Access to music education programs which could help address these issues is limited; in the USA alone, approximately 4 million K-12 students do not have access to music education~\cite{morrison_national_2022}.

To bridge this gap, commercial music tutoring tools have become essential resources. 
Beginner musicians can practice more effectively, and teachers are provided with insights into students' progress~\cite{nart_music_2016,apaydinli_intelligent_2019}. 
The significant demand for such automated solutions is evident, with apps like Yousician~\cite{yousician_ltd_yousician_2024} and Simply Piano~\cite{joytunes_simply_2024} each having over 10 million downloads globally. 

However, Simply Piano and Yousician only identify notes as correct or incorrect, without offering detailed feedback such as missed or extra notes. 
They are also incapable of automatically aligning the user’s performance with a reference score. Instead, Simply Piano and Yousician rely on the user to match their performance with the reference audio. 
Furthermore, their models cannot handle multiple instruments.

The research community has also attempted to provide fine-grained music performance feedback but has had limited success~\cite{benetos_score-informed_2012, wang_identifying_2017}.
A major paradigm of prior work is to temporally align a student's performance with a reference score, and then identify differences.

These alignment-based approaches often fail when there are deviations in the played notes from the score, even if they are minor.
The resulting misalignment of notes leads to inaccurate error detection and ineffective feedback for students. 

\begin{figure}[t]
    \centering
    \includegraphics[width=1\linewidth]{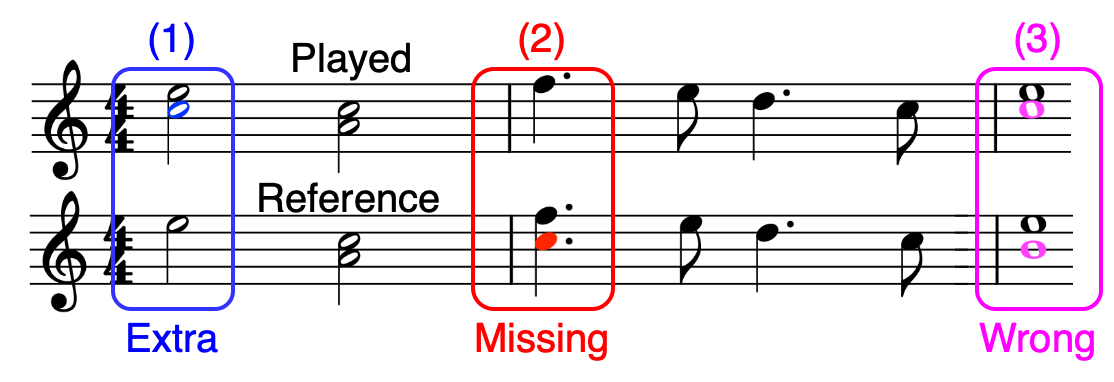}
    \caption{The score on top is the performance transcription and the score below is the reference. 
This paper can detect three types of errors.  
(1) is an extra note. 
“\textit{C}” is played, but it is not expected by the score. 
(2) is a missed note. 
In this figure, “\textit{C}” is not played. 
(3) is a wrong note, which is actually just a missed note and an extra note happening at the same time. 
We expect the player to play a “\textit{C}” but instead a “\textit{B}” is played. 
The model inputs the score and the music student's recorded audio, and labels the notes detected from the recording as “Missed”, “Extra”, or “Correct”.}
\label{fig:error-fig}
\end{figure}

In this paper, we introduce a transformer based-model~\cite{vaswani_attention_2017} called \Polytune to detect musician errors without relying on automatic alignment. 
Our model utilizes an end-to-end trainable approach for error detection.
Specifically, \Polytune takes audio spectrogram pairs as inputs, then outputs an annotated musical score without requiring any manual intervention.
We train \Polytune to annotate notes with a variety of labels beyond ``correct'' or ``incorrect'', as shown in Fig.~\ref{fig:error-fig}.
For training, we require a larger set of training samples than what is provided by existing datasets.
Thus, we introduce an algorithm for synthetically generating errors in existing music datasets, CocoChorales~\cite{wu_chamber_2022} and MAESTRO~\cite{hawthorne_enabling_2018}.
We name the resulting augmented datasets as \textit{CocoChorales-E} and \textit{MAESTRO-E}, respectively.

We evaluate \Polytune and previous works on \textit{CocoChorales-E} and \textit{MAESTRO-E}, which encompass 14 different instruments and a variety of performance errors. 
To evaluate error detection performance, we adapt the transcription F1 score which is commonly used in music transcription tasks~\cite{raffel_transparent_2014}.

We make the following contributions: 
\begin{enumerate}
\item \Polytune achieves \textbf{state-of-the-art} performance in music error detection, with F1 scores of \textbf{95.0\%} for \textit{Correct}, \textbf{49.2\%} for \textit{Missed}, and \textbf{48.0\%} for \textit{Extra}. These results represent improvements of \textbf{+58.0\%}, \textbf{+41.6\%}, and \textbf{+22.1\%}, respectively. 
On average, our method outperforms previous approaches by \textbf{40} percentage points across 14 instruments.
\item We develop an end-to-end training method for audio comparison, eliminating the need for manual alignment.
\item We create a synthetic error generation pipeline for injecting errors into music transcription datasets and create two error detection datasets: \textit{CocoChorales-E} and \textit{MAESTRO-E}.

\end{enumerate}

\section{Background and Related Work}
\label{sec:related_work}

This section explores existing methods for score-informed error detection, advances in token-based music transcription, and the deficiencies of current datasets.
\subsection{Application Context: Music Terminology}
\label{music_terminology}

We define key musical terms used in this paper:
\textbf{Music Score:} A written guide showing notes, rhythms, and instructions for performance. 
\textbf{Music Note:} A symbol indicating the pitch and duration of a sound. 
\textbf{Chord:} Multiple notes played together. 
\textbf{Chordal:} Music with multiple notes at once~\cite{britannica_homophony_2007}
\textbf{MIDI:} A standard that uses symbols to represent musical events like pitch and duration, enabling electronic instruments and computers to communicate. 
\textbf{Track (MIDI Track):} A sequence of MIDI events representing one instrument.
\label{simped}
\subsection{Score-Informed Music Performance Assessment}
\begin{figure}[t]
    \centering
    \includegraphics[width=1\linewidth]{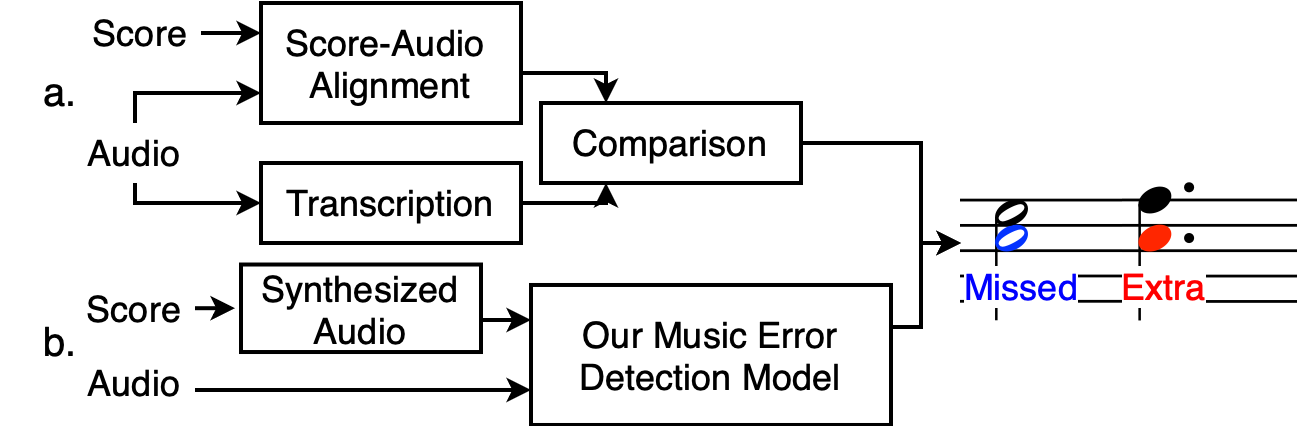}
    \caption{Illustration of differences between (a) previous work and (b) our \Polytune. 
    Our approach simplifies music error detection by using an end-to-end trainable architecture.
    This eliminates the need to explicitly align and compare audio, which is error-prone.}
    \label{fig:compare}
\end{figure}

Music performance assessment requires comparing a player’s rendition against a musical score. 
In \textbf{score-informed} music error detection systems, the score serves as a reference for the detection of errors like missed or extra notes. 
As shown in Fig.~\ref{fig:compare}a, systems by Benetos et al.~\cite{benetos_score-informed_2012} and Wang et al.~\cite{wang_identifying_2017} use Automatic Music Transcription (AMT) to convert audio into musical notation and align it with the score to identify errors like missed or extra notes. 
\begin{figure*}[t]
\centering
\includegraphics[width=0.87\linewidth]{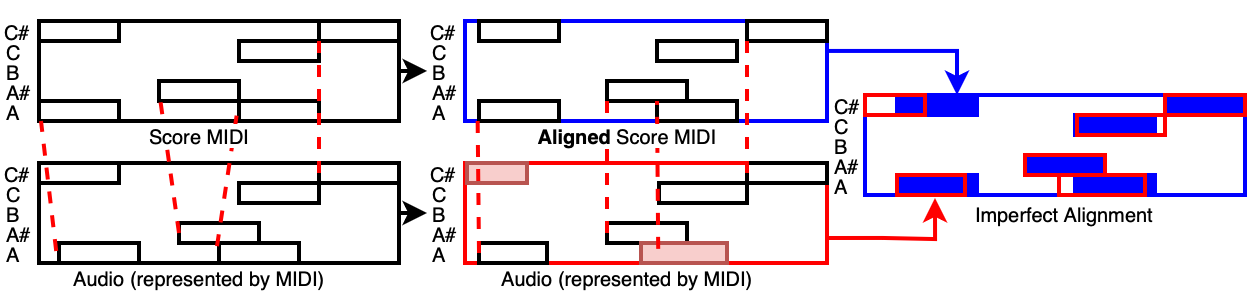}
\caption{
Deficiency of Dynamic Time Warping (DTW): DTW encounters challenges when aligning complex sequences of overlapping notes in music. Specifically, when DTW aligns one note within a group, it often compromises the alignment of other notes. 
This issue arises because DTW attempts to minimize timing differences across the entire sequence. 
Once the MIDI score is aligned with the audio, the algorithm may align one note correctly but misalign others, leading to potential classification errors. 
For example, as seen in the overlay of score and audio, aligning an ``\textit{A}'' note might result in the misalignment of the adjacent ``\textit{C\#}'' note, causing the algorithm to mistakenly classify a correctly played note as a missed ``\textit{C\#}'' and then an extra ``\textit{C\#}''.}
\label{fig:Def}
\end{figure*}
Dynamic Time Warping (DTW)~\cite{sakoe_dynamic_1978} is a widely used technique in music tutoring applications for aligning musical performances with reference scores~\cite{benetos_score-informed_2012,wang_identifying_2017,ewert_score-informed_2016}.
By minimizing temporal differences between two time series, DTW employs dynamic programming to find the optimal alignment.
However, DTW faces challenges when dealing with music that contains overlapping notes, as shown in Fig.~\ref{fig:Def}. 
This forces DTW to compromise, resulting in imperfect alignment or distortion.
Instead, our approach shown in Fig.~\ref{fig:compare}b replaces DTW with a learnable alignment mechanism that does not encounter the same problems as DTW.
We find this leads to more accurate predictions. 

\subsection{Token-Based Automatic Music Transcription (AMT) Methods}
\label{token-based}

Recent AMT advancements are influenced by Natural Language Processing (NLP) techniques, particularly in how outputs are represented. 
In \textbf{token-based} AMT models, the output is represented as sequences of tokens, where each token corresponds to an element of musical information. 
MT3~\cite{gardner_mt3_2022} is a token-based AMT system using a T5~\cite{rael_exploring_2020} transformer to output MIDI-like tokens~\cite{oore_this_2020}. 
These tokens directly represent music and are inspired by the MIDI format.
AMT is treated as a language modeling problem, predicting a sequence of event tokens like time tokens, program tokens, note-on/off tokens, and note tokens.

MT3 uses spectrogram frames as the encoder input, providing context for the autoregressive decoder, which predicts future tokens based on past outputs. This allows the model to capture temporal patterns in music, similar to how NLP models handle text sequences. 
By using token-based representations, MT3 accurately transcribes music. Building on this capability, we adapt MT3 in this paper specifically for error detection.

\subsection{Datasets for Music Error Detection}
\label{data}

Deep learning models require extensive training data. 
Unfortunately, existing datasets for music error detection are limited, with only seven tracks, 15 minutes of audio, and just 40 errors in total~\cite{benetos_score-informed_2012}. 
This scarcity restricts the training of end-to-end models. Training on such a small dataset would cause the model to overfit ~\cite{ying_overview_2019}. 
This highlights the need for more comprehensive datasets to advance research and development in this field.

For AMT, larger datasets like CocoChorales~\cite{wu_chamber_2022} and MAESTRO~\cite{hawthorne_enabling_2018} offer over 1,000 tracks and more than 200 hours of audio.
 Although they are primarily used for transcription tasks, these datasets offer a rich source of musical content.
 In this paper we adapt them to train music error detection models.

\begin{figure*}[ht]
    \centering
    \includegraphics[width=1\linewidth]{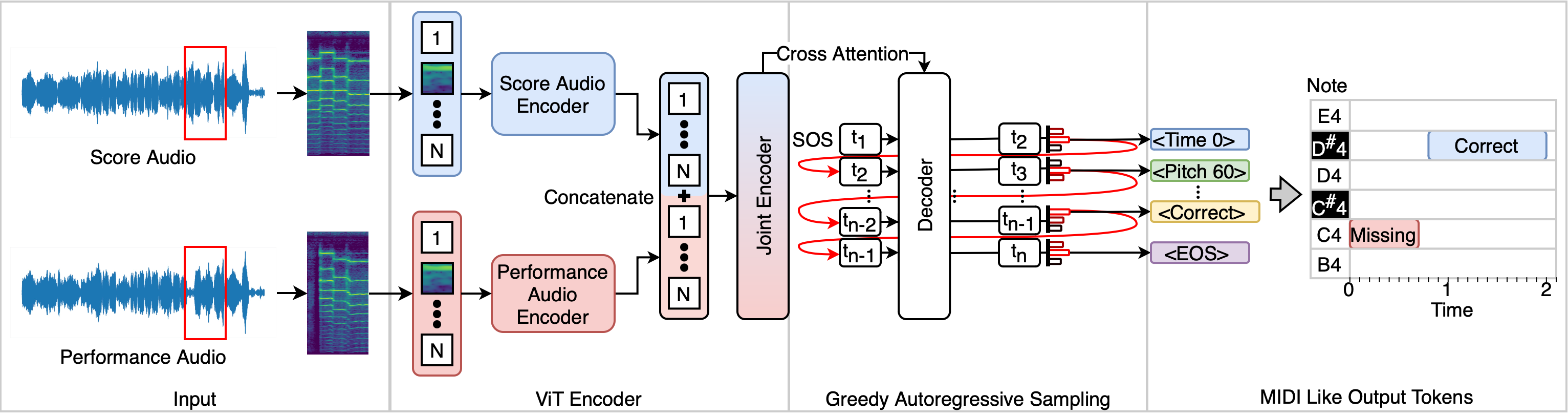}
    \caption{Architecture of \textit{Polytune}. 
    The diagram illustrates the process flow starting with the Score and Performance Audio inputs, each processed through dedicated AST encoders. 
    These encoded features are concatenated and passed through a joint encoder and a decoder with cross-attention for temporal sequencing. 
    The output is generated through greedy autoregressive sampling, providing MIDI-like tokens that classify notes as correct or missing.
    }
    \label{fig:finetune}
\end{figure*}

\section{End-to-end Music Error Detection}
\label{end-to-end}

In this section, we outline key designs of our architecture and training regime.
First, we provide the intuition of our model and describe how it is related to prior work in Sec. \ref{intuition}.
Second, we describe and justify design decisions of our model architecture in  Sec. \ref{error-detection-model}.
Third, we investigate the end-to-end training strategy, and our approach to generating large-scale datasets with synthetic errors in Sec. \ref{input_output}. 
Last, we provide upgraded implementations of~\cite{benetos_score-informed_2012} and~\cite{wang_identifying_2017} to fairly represent older work and contextualize our contributions better in Sec. \ref{baseline}. 
\begin{center}
    \begin{table}[ht]
    \centering
    \label{tab:model_comparison}
    \begin{tabular}{@{}lccccccc@{}}
        \toprule
        \textbf{Model} & \textbf{1} & \textbf{2} & \textbf{3} & \textbf{4} & \textbf{5} & \textbf{6} & \textbf{7} \\
        \midrule
        \textbf{Simply Piano} & \ding{51} & \cellcolor{gray!20}\ding{55} & \cellcolor{gray!20}\ding{55} & \cellcolor{gray!20}\ding{55} & \cellcolor{gray!20}\ding{55} & \cellcolor{gray!20}\ding{55} & \cellcolor{gray!20}\ding{55} \\
        \textbf{Yousician} & \ding{51} & \cellcolor{gray!20}\ding{55} & \cellcolor{gray!20}\ding{55} & \cellcolor{gray!20}\ding{55} & \cellcolor{gray!20}\ding{55} & \cellcolor{gray!20}\ding{55} & \cellcolor{gray!20}\ding{55} \\
        \textbf{Benetos*} & \ding{51} & \ding{51} & \ding{51} & \cellcolor{gray!20}\ding{55} & \cellcolor{gray!20}\ding{55} & \cellcolor{gray!20}\ding{55} & \cellcolor{gray!20}\ding{55} \\
        \textbf{Wang*} & \ding{51} & \ding{51} & \ding{51} & \cellcolor{gray!20}\ding{55} & \cellcolor{gray!20}\ding{55} & \cellcolor{gray!20}\ding{55} & \cellcolor{gray!20}\ding{55} \\
        \textbf{Combined*} & \ding{51} & \ding{51} & \ding{51} & \ding{51} & \cellcolor{gray!20}\ding{55} & \cellcolor{gray!20}\ding{55} & \cellcolor{gray!20}\ding{55} \\
        \midrule
        \textbf{\Polytune*} & \ding{51} & \ding{51} & \ding{51} & \ding{51} & \ding{51} & \ding{51} & \ding{51} \\
        \bottomrule
    \end{tabular}
    \caption{
    Qualitative comparison of models based on key features. 
    We evaluate the models using the following criteria: \textbf{1}: Correct note detection, \textbf{2}: Missed note detection, \textbf{3}: Extra note detection, \textbf{4}: Multi-instrument, \textbf{5}: End-to-end training, \textbf{6}: Automatic alignment, \textbf{7}: No heuristics (Rules or methods based on experience). 
    \textbf{Combined} refers to our baseline model, which reimplements and integrates the methods from Benetos et al. and Wang et al. to enhance performance. 
    Models marked with a star (*) indicate those directly evaluated against in our study.
    }
    \label{tab:compare}
    \end{table}
\end{center}
\subsection{Intuition and Comparison to Prior Work}
\label{intuition}
Our method addresses the main limitations of previous methods, as shown in Tab.~\ref{tab:compare}. 
Benetos et al.~\cite{benetos_score-informed_2012} and Wang et al.~\cite{wang_identifying_2017} rely on time-warping algorithms to align music scores with audio recordings.
These time-warping algorithms suffer from alignment inaccuracies when notes that should be simultaneous fall out of sync.

To overcome these limitations, we introduce a learnable end-to-end model \textit{Polytune}. 
 It processes two unaligned audio sequences—one from the musical score and one from the performance. 
 We design the inputs and outputs to implicitly teach the model alignment, leading to more accurate predictions.
This eliminates the need for complex pre-processing and post-processing steps, such as alignment and comparison, as illustrated in Fig.~\ref{fig:compare}. 
We developed an updated baseline using the score-informed transcription systems of Benetos et al.~\cite{benetos_score-informed_2012} and Wang et al.~\cite{wang_identifying_2017}, incorporating the MT3 model and Dynamic Time Warping (DTW) to reflect current best practices (Sec. \ref{baseline}).

\subsection{Music Error Detection Model}
\label{error-detection-model}

We show how our model architecture can be designed to implicitly learn better sequence alignment in latent representation, which is crucial for accurately detecting errors in music performances. 
Since transcription plays a key role in existing methods for music error detection, we base our approach on MT3. 
Specifically, we frame music error detection as a sequence-to-sequence task, where the input consists of audio spectrogram frames and the output is represented by customized tokens for error detection.

Hawthorne et al. treats music transcription as a sequence-to-sequence modeling task, using audio spectrogram frames as inputs and MIDI-like tokens as outputs~\cite{hawthorne_sequence--sequence_2021}. 
Building on this foundation, MT3~\cite{gardner_mt3_2022} advanced the piano transcription model by generalizing it to transcribe different instruments. 
Therefore, we base our model’s decoder on MT3, a well-studied state-of-the-art Automatic Music Transcription (AMT) method.

The architecture of \Polytune is detailed in Fig.~\ref{fig:finetune}. 
The design fuses two inputs: the audio from the musical score and the player’s corresponding performance. 
These are encoded into a joint encoding space.
We treat our model as multi-modal, utilizing modality-specific encoders similar to those in~\cite{gong_contrastive_2023,akbari_vatt_2021}. 
Our reasoning is that although both inputs are audio spectrograms, the score and performance audio serve distinct roles in music error detection. 

We fuse two Audio Spectrogram Transformer~\cite{gong_ast_2021} (AST) encoders with a joint encoder.
Then, we use a standard T5 decoder, with greedy autoregressive decoding to output sequences of tokens, described in Sec.~\ref{input_output}. To bridge the fused AST encoders and the T5 decoder, we employ a linear projection layer to transform the embedding size from AST’s 768 to T5’s 512. The encoder outputs condition the decoder via a cross attention mechanism.

\subsection{End-to-End Training of the Music Error Detection Model}
\label{input_output}

\subsubsection{Input Data}  
We use audio inputs synthesized from musical scores rather than symbolic representations like MIDI tokenizations. 
This approach has several advantages:

\begin{enumerate}

\item \textbf{Generalization:}  
    Audio inputs generalize across instruments and capture their full sound, including characteristics like vibrato (violin), trills (flute), and pitch bending (guitar), which MIDI tokenizations fail to represent. MIDI tokenizations are designed primarily for piano. Their design prohibits them from representing non-fixed-pitch instruments or capturing instrument-specific nuances (e.g., see~\cite{hsiao_compound_2021, huang_pop_2020}).

\item \textbf{Extensibility and Expressiveness:}  
    Audio preserves absolute timing and dynamics, enabling the detection of new error types, such as timing or dynamics errors, without requiring changes to the input representation.
    In contrast, MIDI tokenizations reduce sequence length to address the $O(N^2)$ computational cost of transformers but sacrifice expressiveness and timing precision. For instance, the Octuple tokenizer achieves a fourfold reduction in sequence length compared to REMI~\cite{huang_pop_2020}, but introduces errors in complex time signatures~\cite{fradet_miditok_2021}.

\item \textbf{Comparability:}  
Audio inputs allow users to compare performances against any pre-recorded audio, enabling analysis in contexts where scores are unavailable or nonexistent, such as impromptu jazz or folk music.

\item \textbf{Simplicity in Design:}  
Using audio for both inputs simplifies the model by reusing the same feature extractor design for both encoders. We conjecture that using a single input format for the music score and performance enables the model to learn implicit alignment more effectively. This avoids the additional capacity required to process and compare different modalities. By using the same input format, we eliminate unnecessary complexity from the task.
    
\end{enumerate}

Fig.~\ref{fig:finetune} illustrates our data flow.  
We divide the student’s recorded audio and the synthesized score audio into non-overlapping segments. 
The segment length is 2.145 seconds.
Each segment undergoes a short-time Fourier transform, producing spectrogram frames. 
Finally, we tokenize each spectrogram frame using the ViT patch embedding approach~\cite{dosovitskiy_image_2021} and feed the resulting 512 patches into our ViT encoder.

\begin{table}[ht]
\centering
\begin{tabular}{p{1.2cm}p{4.8cm}r} 
\toprule
\textbf{Name} & \textbf{Description} & \textbf{Values} \\
\midrule
\textbf{SOS} & Start of sequence & 1 \\
\midrule
\textbf{End Tie} & Defines non-active notes from previous segments & 1 \\
\midrule
\textbf{Time} & Specifies note timing in a segment & 205 \\
\midrule
\textbf{Label} & Note is extra, missed, or correct & 3 \\
\midrule
\textbf{On/Off} & Note is played or not & 2 \\
\midrule
\textbf{Note} & MIDI Pitch & 128 \\
\midrule
\textbf{EOS} & End of sequence & 1 \\
\bottomrule
\end{tabular}
\caption{Vocabulary for MIDI note events, adapted from \cite{gardner_mt3_2022}. 
Compared with their vocabulary, we add additional note labels ``extra'' and ``missed''. 
}
\label{tab:vocab}
\end{table}

\subsubsection{Output Data}
We output a sequence of MIDI-like tokens to represent the music scores with errors annotated.
The complete token vocabulary is summarized in Tab.~\ref{tab:vocab}.
The vocabulary is similar to~\cite{gardner_mt3_2022, hawthorne_sequence--sequence_2021} with the following differences:

\begin{itemize}
    \item We do not specify instrument because we assume in a music tutoring context musicians would typically play a single instrument. 
    Our error detection model is instrument agnostic, because we use the same output formats for errors of different instruments.
    \item We add “Label” tokens to annotate each note with one of the error categories defined in Sec.~\ref{intro}.
\end{itemize}

While Compound Word~\cite{hsiao_compound_2021} and Revamped MIDI (REMI)~\cite{huang_pop_2020} tokenizations offer efficient sequence lengths compared to Midi-like tokenization, they come with drawbacks. 
These methods trade off fine-grained timing accuracy for sequence length by using bars and beats to represent time.
MIDI-like tokenization for output captures timing more precisely, and leads to more accurate predictions.

\subsubsection{Generating Datasets}
A large number of labeled musician errors are required for end-to-end training. 
However, no large-scale datasets exist for music error detection. 
The only existing dataset by Benetos et al.~\cite{benetos_score-informed_2012} has just 7 tracks.

In order to address this gap, we create two datasets, \textit{MAESTRO-E} and \textit{CocoChorales-E}, that have 1000+ samples for each instrument.  
\textit{MAESTRO-E} contains over 200 hours of audio,
with 1,000+ tracks focused on piano, including 200k note
and timing errors. 
\textit{CocoChorales-E} includes 300+ hours of audio across 10k+ tracks and 13 instruments, featuring over 25k note and timing errors. 
The dataset from Benetos et al.~\cite{benetos_score-informed_2012} is much smaller, with only 15 minutes of audio, 7 tracks, and 40 note errors. 
We use MIDI samples from CocoChorales and MAESTRO augmented with common performance errors such as wrong note, missed note, and extra note. 
Additionally, we synthesize audio using MIDI-DDSP~\cite{wu_midi-ddsp_2022} for each of the augmented samples. 
We define training labels by splitting our dataset into three MIDI files for the classes: \textit{Correct}, \textit{Missed}, and \textit{Extra} as defined in Sec. \ref{intro}.

\begin{table*}[t]
\centering
\setlength{\tabcolsep}{8pt} 
\begin{tabular}{llrrrrrr}
\toprule
\textbf{Category} & \textbf{Method} & \multicolumn{3}{c}{\textbf{\textit{MAESTRO-E}}} & \multicolumn{3}{c}{\textbf{\textit{CocoChorales-E}}} \\
\cmidrule (lr){3-5} \cmidrule (lr){6-8}
& & \textbf{Precision} & \textbf{Recall} & \textbf{F1 Score} & \textbf{Precision} & \textbf{Recall} & \textbf{F1 Score} \\
\midrule
\multirow{2}{*}{Correct}
& \Polytune & \textbf{96.9\%} & \textbf{84.1\%} & \textbf{90.1\%} & \textbf{96.2\%} & \textbf{94.8\%} & \textbf{95.4\%} \\
& Baseline & 46.8\% & 40.7\% & 43.5\% & 42.3\% & 33.1\% & 36.7\% \\
\midrule
\multirow{2}{*}{Missed}
& \Polytune & \textbf{30.7\%} & \textbf{26.3\%} & \textbf{26.8\%} & \textbf{51.6\%} & \textbf{55.1\%} & \textbf{51.3\%} \\
& Baseline & 3.9\% & 24.1\% & 6.6\% & 5.3\% & 17.3\% & 7.7\% \\
\midrule
\multirow{2}{*}{Extra}
& \Polytune & \textbf{70.5\%} & 76.3\% & \textbf{72.0\%} & \textbf{47.1\%} & \textbf{48.5\%} & \textbf{46.8\%} \\
& Baseline & 26.3\% & \textbf{87.9\%} & 39.9\% & 16.1\% & 52.6\% & 23.5\% \\
\midrule
\multirow{2}{*}{Average}
& \Polytune & \textbf{64.9\%} & \textbf{62.2\%} & \textbf{62.9\%} & \textbf{65.0\%} & \textbf{66.1\%} & \textbf{64.5\%} \\
& Baseline & 25.6\% & 50.9\% & 30.0\% & 21.2\% & 34.6\% & 22.7\% \\
\bottomrule
\end{tabular}
\caption{Comparison of \Polytune and Baseline across Error Types in Two Datasets (\textit{MAESTRO-E} and \textit{CocoChorales-E}). 
Boldface indicates the best performance in each category. 
Our method demonstrates superior performance in most cases, except for a higher recall for extra notes in the Baseline. 
This is explained in Sec.~\ref{sec:quantitative}.}
\label{tab:comparison_datasets}
\end{table*}

\begin{table*}[t]
    \centering
    \setlength{\tabcolsep}{4pt} 
    \begin{tabular}{lcccccc}
        \toprule
        \textbf{Instrument} & \multicolumn{2}{c}{\textbf{Correct}} & \multicolumn{2}{c}{\textbf{Missed}} & \multicolumn{2}{c}{\textbf{Extra}} \\
        \cmidrule(lr){2-3} \cmidrule(lr){4-5} \cmidrule(lr){6-7}
        & \textbf{\Polytune} & \textbf{Baseline} & \textbf{\Polytune} & \textbf{Baseline} & \textbf{\Polytune} & \textbf{Baseline} \\
        \toprule
        Average & \textbf{95.0\%} & 37.0\% & \textbf{49.2\%} & 7.6\% & \textbf{48.0\%} & 25.9\% \\
        \midrule
        Piano & \textbf{90.1\%} & 43.5\% & \textbf{26.8\%} & 6.6\% & \textbf{72.0\%} & 39.9\% \\
        Flute & \textbf{96.0\%} & 38.9\% & \textbf{56.0\%} & 7.2\% & \textbf{52.0\%} & 26.6\% \\
        Clarinet & \textbf{95.6\%} & 38.3\% & \textbf{49.7\%} & 6.7\% & \textbf{46.6\%} & 24.1\% \\
        Oboe & \textbf{96.3\%} & 33.4\% & \textbf{58.4\%} & 6.7\% & \textbf{48.1\%} & 25.9\% \\
        Bassoon & \textbf{94.4\%} & 34.7\% & \textbf{48.9\%} & 6.4\% & \textbf{41.7\%} & 17.1\% \\
        Violin & \textbf{95.5\%} & 36.1\% & \textbf{57.1\%} & 7.5\% & \textbf{48.8\%} & 27.3\% \\
        Viola & \textbf{95.1\%} & 36.1\% & \textbf{46.9\%} & 5.9\% & \textbf{47.7\%} & 26.1\% \\
        Cello & \textbf{94.9\%} & 37.5\% & \textbf{42.7\%} & 6.9\% & \textbf{46.8\%} & 21.7\% \\
        Trumpet & \textbf{96.3\%} & 37.8\% & \textbf{58.7\%} & 8.8\% & \textbf{53.6\%} & 26.6\% \\
        French Horn & \textbf{96.1\%} & 38.4\% & \textbf{53.9\%} & 5.9\% & \textbf{43.2\%} & 23.7\% \\
        Tuba & \textbf{95.2\%} & 37.3\% & \textbf{45.4\%} & 8.1\% & \textbf{45.6\%} & 17.8\% \\
        Trombone & \textbf{94.8\%} & 35.0\% & \textbf{50.4\%} & 7.1\% & \textbf{44.8\%} & 21.7\% \\
        Contrabass & \textbf{94.2\%} & 35.7\% & \textbf{42.0\%} & 8.9\% & \textbf{38.6\%} & 19.9\% \\
        Tenor Sax & \textbf{95.7\%} & 39.7\% & \textbf{56.2\%} & 14.2\% & \textbf{45.7\%} & 25.1\% \\
        \bottomrule
    \end{tabular}
    \caption{Error Detection F1 Score Comparison by Instrument and Error Category. 
    Piano is from \textit{MAESTRO-E} and the rest are from \textit{CocoChorales-E}.}
    \label{tab:instrument_f1_score_comparison}
\end{table*}

The algorithm for generating MIDI errors is detailed in Alg.~\ref{alg:midi_error_generation}.
\begin{algorithm}
\begin{algorithmic}[1]
\Require All notes in MIDI track $A$, error rate $\lambda$, offset distributions $P$, $Q$

    \State Select notes from $A$ to augment with probability $\lambda$
    \For{each note selected}
        \State $\text{err\_type} \gets \text{rand}(\, \{\text{miss, PC, TS, EN}\} \,)$
        \If{$\text{err\_type} = \text{miss}$} 
            \State Remove note;
        \ElsIf{$\text{err\_type} = \text{PC}$}
            \State $\epsilon_p$ $\gets$ sample($P$)
            \State Offset pitch by $\epsilon$;
        \ElsIf{$\text{err\_type} = \text{TS}$} 
            \State $\epsilon_t$ $\gets$ sample($Q$)
            \State Offset time by $\epsilon$;
        \ElsIf{$\text{err\_type} = \text{EN}$} 
            \State $\epsilon_p$ $\gets$ sample($P$)
            \State $\epsilon_t$ $\gets$ sample($Q$)
            \State Insert note with time offset $\epsilon_t$ and pitch offset $\epsilon_p$;
        \EndIf
    \EndFor
\end{algorithmic}
\caption{MIDI Error Generation Algorithm. 
This algorithm introduces errors into MIDI files. 
Abbreviations: \textit{PC} (pitch change), \textit{TS} (timing shift), \textit{EN} (extra note).}
\label{alg:midi_error_generation}
\end{algorithm}
We introduce errors into each MIDI file by selecting notes based on a Poisson distribution with a mean rate parameter $\lambda$, where $\lambda$ is sampled from a uniform distribution $U(0.1, 0.4)$ and applying an error type.
Then, the randomly selected notes are assigned an error type, and their time and pitch are augmented accordingly.
Offset magnitudes for time and pitch are sampled from two truncated normal distribution distributions, $P$ and $Q$, with mean 0 and standard deviation of 1 and 0.02, respectively.
Note that $P$ and $Q$ can have different parametrizations.
We believe our choice of truncated normal distributions simulate reasonable performance errors in pitch and timing~\cite{trommershauser_optimal_2005, tibshirani_statistician_2011}.

\subsection{Baseline}
\label{baseline}
We choose Benetos et al. and Wang et al. as our baseline for score-informed error detection~\cite{benetos_score-informed_2012, wang_identifying_2017}.
They are the most relevant prior works to score informed error detection.
\textit{We created upgraded, open-source versions of these works to enable fairer comparisons.}

In our re-implementation of Benetos et al. and Wang et al., we retain the foundational concepts of each approach but incorporate changes to each component of the transcription pipeline to reflect advances in automatic music transcription (AMT).
Specifically, we replace the original non-negative matrix factorization-based transcription with the state-of-the-art MT3 model.
Additionally, we use Dynamic Time Warping (DTW) instead of the less accurate Windowed Time Warping (WTW). 
Our updated baseline shows comparable performance and also includes the ability to detect multiple instruments.

\begin{figure}[!ht]
    \centering
    \includegraphics[width=0.92\linewidth]{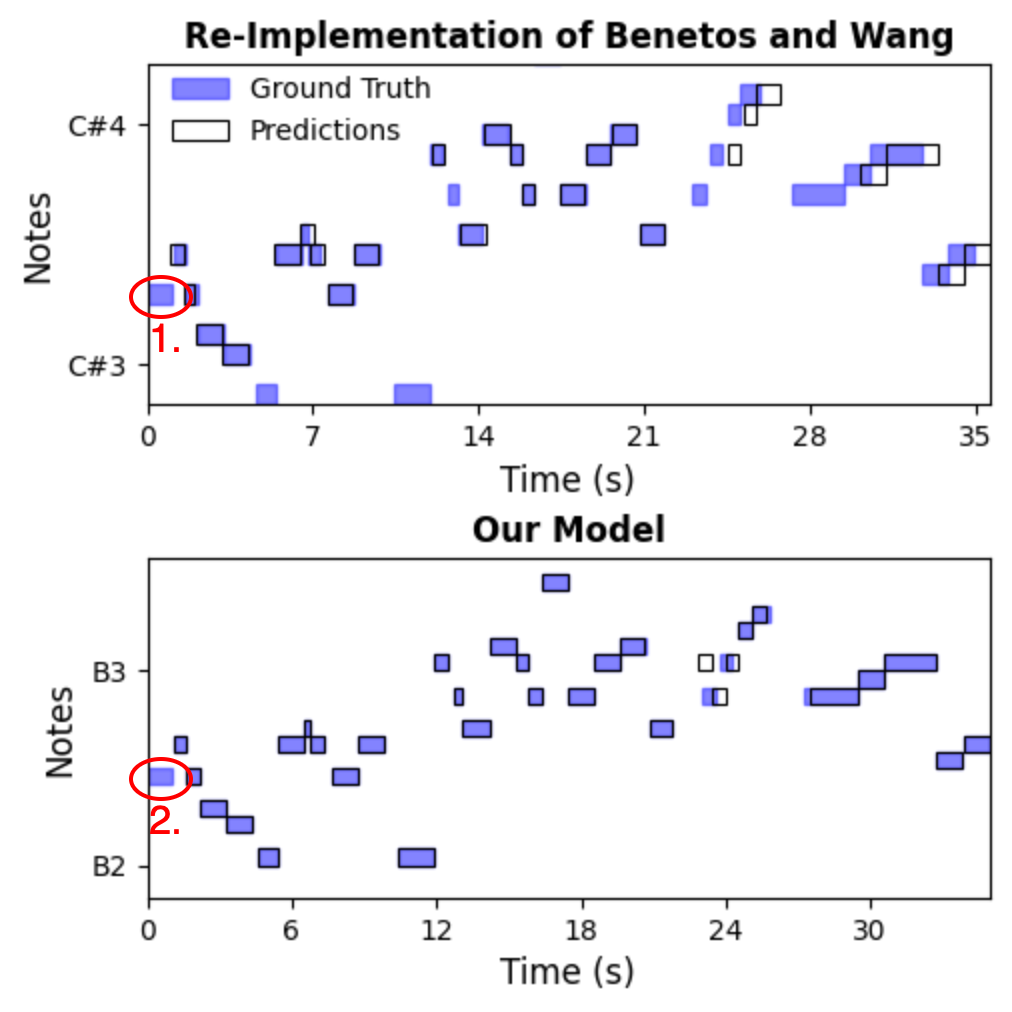} 
    \caption{Qualitative Comparison of MIDI Note Events: This figure shows the ``correct category'' note events detected by each model for a track in the \textit{CocoChorales-E} dataset. 
    Ground truth notes are filled blue rectangles, and model predictions are black-outlined rectangles. 
    Music note 1 is caused by a transcription error by MT3 and a similar error occurs with Music note 2. 
    Overall, \Polytune better matches the ground truth and has fewer false detections than our Benetos and Wang re-implementation.
    }
    \label{fig:qualitative-comparison-fig}
\end{figure}

\section{Results and Discussion}
In this section, we compare \Polytune with prior work across our datasets \textit{MAESTRO-E} and \textit{CocoChorales-E}.
First, we describe the environment used to train and test our approach in Sec. \ref{sec:experimental}.
Second, we present the performance of our model and the baseline in Sections \ref{sec:quantitative} and \ref{sec:qualitative}.
Third, insight regarding the training of \Polytune is discussed in Sec. \ref{sec:ssl}.

\subsection{Experiment Setup}
\label{sec:experimental}
We used Pytorch 2.3.0 and Huggingface Transformers 4.40.1 for model design and training.
The mir\_eval package is used for evaluating Error Detection F1 scores.

All models were trained on an NVIDIA A100-80GB GPU running a Linux operating system. 
The datasets introduced in this work, \textit{MAESTRO-E} and \textit{CocoChorales-E}, were generated using AMD EPYC 7713 3.0 GHz CPUs. Dataset generation took approximately 12 hours per dataset, with \textit{MAESTRO-E} using 1 CPU and \textit{CocoChorales-E} using 256 CPUs in parallel.

For model evaluation, we define the Error Detection F1 score for each error type—Extra (E), Missed (M), and Correct (C)—by isolating the corresponding notes and applying the transcription onset F1 metric from \texttt{mir\_eval}~\cite{raffel_transparent_2014}. 
All results are based on a combined test set of 4401 tracks.

\subsection{Quantitative Evaluation}
\label{sec:quantitative}
We present a comparison of our method against the baseline across different categories for Error F1, precision, and recall. 
As shown in Tab.~\ref{tab:comparison_datasets}, our method generally outperforms the baseline derived from~\cite{benetos_score-informed_2012, wang_identifying_2017}, except for a higher recall of 87.9\% in extra notes for \textit{MAESTRO-E} by the baseline. 
This discrepancy arises because extra and missing notes disrupt DTW alignment, leading the baseline’s heuristics to mark many unmatched notes as extra. 
However, this misalignment also causes correct notes to be misclassified as extra, resulting in high recall but low precision, particularly in \textit{MAESTRO-E}. 
Overall, for extra notes in \textit{MAESTRO-E}, \Polytune still outperforms the baseline by 32.1 F1 percentage points. 
A similar pattern is seen in missed notes, where the baseline shows higher recall but lower precision. 
Additionally, we compared Error Detection F1 scores by instrument (Tab.~\ref{tab:instrument_f1_score_comparison}). 
Our findings indicate that explicit alignment is not essential for effective music error detection.

\subsection{Qualitative Evaluation}
\label{sec:qualitative}

Fig.~\ref{fig:qualitative-comparison-fig} compares \Polytune error detection results with the baseline.
Alignment errors are visible as offsets between the outlined and colored boxes. 
The baseline model often fails to detect missing and extra notes due to DTW alignment issues.
Meanwhile, \Polytune does not make the same mistakes and achieves greater precision.
 
The baseline model’s alignment errors stem from DTW struggling to align examples containing mistakes with the reference score. 
These alignment errors are the main reason for the baseline's false detections.
However, \Polytune also struggles with notes that transcription models find challenging. 
See Fig.~\ref{fig:qualitative-comparison-fig} for an example where MT3 misdetects Note 1, and \Polytune makes a similar error with Note 2. 

\subsection{Insights on Training Music Error Detection Models}
\label{sec:ssl}
In our work, we identify two key insights for training our music error detection models. 
The first insight involves addressing the class imbalance in our datasets, which we tackle using a weighted cross-entropy loss. 
The second insight relates to the limited benefits of self-supervised pretraining for this task, causing us to prioritize direct supervised learning instead.

First, the \textit{MAESTRO-E} and \textit{CocoChorales-E} datasets are class imbalanced, with \textit{MAESTRO-E} having an average error-to-correct ratio of 1:9 and \textit{CocoChorales-E} having a ratio of 1:4. 
To address this imbalance, we use a weighted cross-entropy loss, as shown in Equation~\ref{class_imbalance}.

\begin{equation}
\mathcal{L} = \frac{1}{N} \sum_{i=1}^{N} \left[ \alpha(y_i) \cdot \text{CE}(y_i, \hat{y}_i) \right]
\label{class_imbalance}
\end{equation}

Equation~\ref{class_imbalance} defines the weighted cross-entropy loss $\mathcal{L}$, averaged over $N$ tokens. 
$\text{CE}(y_i, \hat{y}_i)$ is the cross-entropy between true label $y_i$ and prediction $\hat{y}_i$, weighted by a function $\alpha(y_i)$. 
For our training, $\alpha(y_i)$ is 10 when $y_i$ is an error token and 1 when it is not. 

Second, we explored the effectiveness of self-supervised pretraining in improving error detection. 
Previous works, such as those by Huang et al.~\cite{huang_masked_2022} and Gong et al.~\cite{gong_contrastive_2023}, utilized this approach for audio classification and retrieval. 
However, we found no benefit to the final Error Detection F1 scores in our case. 
Considering the significant hardware and data requirements for pretraining, we decided to train \Polytune with direct supervised learning instead. 
Preliminary results are provided in the Appendix, along with potential explanations for these findings.

\section{Limitations and Future Work}
\label{sec:limitations}

\Polytune under-performs on missed notes in homo-phonic music (see Tab.~\ref{tab:comparison_datasets}). 
We conjecture that this is due to the challenges of representing chordal textures in spectrogram form.
Furthermore, the use of only one synthesizer per instrument in our dataset may restrict \textit{Polytune}’s ability to generalize to other datasets or real-world performances with diverse timbres. 
This limitation could be addressed by incorporating a wider variety of synthesized timbres, which would enhance the model’s robustness and adaptability.
Lastly, our model’s vocabulary, designed around the 12-tone equal temperament system, may limit equitable music education for non-Western cultures; future work can expand to more inclusive vocabularies.

\section{Conclusion}
\label{sec:conclusion}
This paper proposes \textit{Polytune}, an end-to-end trainable transformer model for music error detection. 
\Polytune outperforms previous methods without automatic alignment methods. Additionally, we introduce synthetic data generation to improve error-detection performance. 
We generate synthetic errors for 14 instruments to train and evaluate \Polytune and the baseline method. 
\Polytune consistently outperforms the baseline, achieving a 64.1\% Error Detection F1. 
This represents a 40 percentage point improvement across 14 instruments. 
These findings demonstrate the effectiveness of our end-to-end transformer-based approach and the significant role of synthetic data in advancing music error detection.

\end{document}